\documentclass[reprint,aps,prd,twocolumn,superscriptaddress,nofootinbib,showkeys]{revtex4-2}
\usepackage[utf8]{inputenc}
\usepackage[colorlinks=true, linkcolor=blue, citecolor=blue, urlcolor=blue]{hyperref}
\usepackage{amsmath, amssymb, amsfonts, bm, graphicx, microtype, orcidlink, esvect, url, booktabs, amsthm, adjustbox, placeins, csquotes, changes, xspace, multirow, bm}
\usepackage[normalem]{ulem}

\newcommand{\m}{M\textsubscript{\(\odot\)}\xspace}

\begin{document}

\title{Search for Sub-Solar Mass Binaries in the First Part of LIGO’s Fourth Observing Run}

\author{Keisi Kacanja\orcidlink{0009-0004-9167-7769}}
\email[Corresponding author: ]{kkacanja@syr.edu}
\author{Kanchan Soni\orcidlink{0000-0001-8051-7883}}
\author{Aleyna Akyüz\orcidlink{0000-0001-7248-951X}}
\author{Alexander H. Nitz\orcidlink{0000-0002-1850-4587}}
\affiliation{Department of Physics, Syracuse University, Crouse Dr, Syracuse, NY 13210}

\date{\today} 

\begin{abstract} 
We report the first results of a sub-solar mass compact binary search using the data from the first part of the fourth observing run (O4a) of the Advanced LIGO detectors. Sub-solar mass neutron stars and black holes are not expected to form via standard stellar evolution, and their observation would signify a new class of astrophysical objects or the discovery of a component of dark matter. Our search covers binaries with primary masses 0.1 to 2 $M_\odot$ and secondary masses 0.1 to 1 $M_\odot$. We explicitly incorporate tidal effects up to $7\times10^5$ for extremely low mass neutron stars. Due to the recent development of efficient ratio filter de-chirping frameworks, this search consisting of 25 million templates is now computationally feasible. No statistically significant candidates are identified. We place a $90\%$ confidence upper limit on the merger rate $\mathcal{R}_{90}$  for sub-solar mass black holes to be $< 2.5\times10^4\,\textrm{Gpc}^{-3} \textrm{yr}^{-1}$ for a chirp mass of 0.2 $M_\odot$. We place the first constraints for binary neutron stars with tidal deformabilities up to $\sim 7\times10^5$ and improve the merger rate estimate by a factor $\sim 3$ in comparison to previous O3 tidal searches for tidal deformabilities $< 10^4$. The advanced sensitivity of the O4a run enables an improvement in the sub-solar mass black hole merger rate limits by more than $2 \times$ over the previous three observing runs (O1-O3) combined. We constrain the effective local dark matter fraction to be  $\tilde{f}_\textrm{PBH}<0.5\%$ for 0.4 $M_{\textrm{PBH}}$, approximately 1.8 times lower than the previous O1-O3 constraints. Given our model assumptions, our local dark matter fraction constraints are 2-10 times lower than the OGLE microlensing survey for $M_{\textrm{PBH}}\ge0.25$.

\end{abstract}


\maketitle

\section{Introduction} \label{sec:intro}

In 2015, the Advanced LIGO detectors in Hanford and Livingston began operations \citep{advancedligo}, and made the first direct detection of the gravitational wave GW150914 resulting from the merger of a binary black hole system (BBH) on September 14, 2015 \citep{2016PhRvL.116f1102A}. This was followed by the first binary neutron star (BNS) merger, GW170817 \citep{2017PhRvL.119p1101A,Alexander:2017aly}, observed through both gravitational waves and electromagnetic counterparts \citep{2021ApJ...918L..29R,2017Sci...358.1556C,2017ApJ...848L..30P,2019ApJ...887L..24M,2019ApJ...887L..21R,LIGOScientific:2017ync,Valenti_2017}, and two neutron star–black hole mergers detected in January 2020 \citep{2021ApJ...915L...5A}. To date, the LIGO-VIRGO-KAGRA (LVK)~\citep{advancedligo,advancedvirgo,kagra} observatory network has detected $\sim 300$ sources, predominantly BBHs, four confident neutron star black hole binaries, and two BNS detections \citep{1gwtc,2gwtc,3gwtc,4gwtc}. At present, the majority of gravitational-wave observations have been consistent with current stellar evolution predictions with component masses above $1\,M_{\odot}$\citep{4gwtcpop} for both neutron stars and black holes. To this day, no confident sub-solar mass (SSM) compact binary has been observed. Recent low-significance alerts have highlighted potential SSM candidates, such as S251112cm that is estimated to have a false-alarm rate of one in four years~\citep{S251112cmgcn}. Previous studies and searches have also reported potential SSM candidate events, motivating continued interest in this mass range \citep{2022NatAs...6.1444D,Yuan:2024yyo,ligosubsolar,nitzsubsolar1,nitzsubsolar2,2025ApJ...984...61K}. The absence of a confident observation may reflect the limited detector sensitivity to low-mass systems or the rarity of these systems.

Observing SSM compact objects would motivate the exploration of non-standard formation channels, new physics, and reveal unique compact objects such as Primordial Black Holes (PBHs) \citep{2024PhRvD.109l4063C}. PBHs are hypothesized to form in the very early universe from the direct collapse of over-dense regions in a largely pristine environment \citep{darkmatterreview,PBHreview}. Because they form in regions with little baryonic matter, PBHs could retain surrounding dark matter (DM), potentially providing a direct measurement of DM if observed \citep{darkmatterreview}. 

It is currently unknown whether DM consists of subatomic particles, such as WIMPs or axions, or macroscopic objects like PBHs \citep{darkmatterreview}. In the latter case, PBHs themselves could constitute a fraction of DM \citep{PBHreview,darkmatterreview,PBHasDM}. Alternatively, if DM is a particle, another theory suggests that PBH could host a halo of these particles, and interactions with a binary companion could reveal properties of the surrounding matter that can be detected by sensitive ground based interferometers \citep{2023PhRvD.107h3006C,Boudaud:2021irr}, though this scenario is challenging to observe.  

PBHs could span a wide range of masses, which depends on their formation epoch \citep{greenPBHreview}. PBHs formed during the cosmic QCD epoch are expected to have characteristic masses in the SSM range \citep{Byrnes:2018clq, Clesse:2020ghq, Jedamzik:2020omx, Jedamzik:2020ypm, Juan:2022mir}. PBHs lighter than $10^{-16}~\m$ would have evaporated due to Hawking radiation and are unlikely to survive to the present day \citep{greenPBHreview}. At intermediate masses, microlensing, Kepler transient surveys, and Cosmic Microwave background measurements have placed stringent limits on the abundance on PBHs \citep{Macho:2000nvd,EROS-2:2006ryy,Griest:2013aaa,Niikura:2019kqi,Mroz:2024mse,PBHbounds}. Gravitational wave searches provide additional constraints, limiting the PBHs to contribute a small fraction of the total DM density. If observed, SSM black holes would not only be decisive for the existence of PBHs but could also point to other exotic formation scenarios, such as dark-matter–driven collapse producing SSM dark black holes \citep{Shandera:2018xkn,Dasgupta:2020mqg,Singh:2020wiq,Oguri:2017ock,Croon:2020ouk}.

Previous direct SSM black hole searches have placed meaningful constraints on this abundance of DM for the $\sim10^{-2}$ \m  to $\sim10^2$ \m mass window \citep{greenPBHreview,ligosubsolar,nitzsubsolar,nitzsubsolar1,nitzsubsolar2,nitzsubsolarhighq}.
Prior searches placed constraints on the DM fraction to be $<2\%$ for 0.2 \m  PBH \citep{nitzsubsolar}.
The recent release of significantly improved O4a data \cite{2025PhRvD.111f2002C} provides an opportunity to further constrain the upper limits on the local DM fraction.

Another potential SSM compact object could be a neutron star. Theoretically, the minimum stable mass of a neutron star is predicted to be $\sim 0.1\, M_{\odot}$ \citep{1983bhwd.book.....S}. While the formation channels capable of producing neutron stars with masses $<1\,M_{\odot}$ are expected to be extremely limited, one proposed mechanism involves gravitational instability or fragmentation of accretion disks formed from the collapse of a rapidly rotating massive star \citep{Piro:2006ja,Metzger_2024}. Three-dimensional hydrodynamical simulations of neutrino-cooled disks show that, under sufficiently high cooling rates, the outer disk can fragment into neutron-rich, self-gravitating clumps with masses between $0.01$–$1 ~M_{\odot}$ \citep{Chen:2025uwd}. Many of these clumps are expected to exceed the critical mass for gravitational collapse and may form SSM neutron stars. \citep{Metzger_2024,Chen:2025uwd}. If such objects remain in bound systems, they could serve as potential sources detectable by ground-based gravitational-wave observatories.

Recent multi-messenger observations further motivate targeted searches for SSM binaries. In particular, the subthreshold gravitational wave trigger S250818k \citep{S250818k} reported by the LVK collaboration represents a potential candidate with at least one component mass estimated to be below $1 \, \m$. Additionally, an optical transient, ZTF25abjmnps (AT2025ulz), was identified to lie within the 786-949 $\textrm{deg}^2$ localization region of S250818k \citep{2025GCN.41451....1S,GCN42032}, raising the possibility of a multi-messenger association, although this association is not confirmed~\citep{ZTF25abjmnps}. If astrophysical in origin, this event may constitute the first tentative observational evidence for a SSM neutron star merger and offer support for the fragmentation scenario as a formation channel \citep{Piro:2006ja,Metzger_2024,Chen:2025uwd,ZTF25abjmnps}. However, the combination of supernova-like spectral features, a kilonova-like early light curve, and the absence of radio and X-ray emission leaves its physical origin ambiguous \citep{ZTF25abjmnps}. While the association remains tentative, S250818k underscores the need for gravitational-wave searches that are sensitive to SSM neutron stars.

The observation of a SSM neutron star would uniquely probe the dense matter equation of state (EOS) \citep{2007Ap&SS.308..371L,2021ApJ...918L..29R,2023APS..APRM03003D,Rutherford:2024srk,Raaijmakers_2021}. Unlike black holes, neutron stars undergo tidal deformation in a binary, leaving a measurable imprint on the gravitational waveform quantified by the effective tidal deformability $\tilde{\Lambda}$ \citep{2008PhRvD..77b1502F,2010PhRvD..81l3016H}. Current EOS models predict that tidal deformability can scale from $\mathcal{O}(10^3)$ for a $1\,M_{\odot}$ star up to $\mathcal{O}(10^8)$ for a $0.1 ~M_{\odot}$ star \citep{2020CQGra..37d5006A,1997A&A...328..274B,MULLER1996508,DANIELEWICZ200936,PhysRevC.72.014310,PhysRevC.53.740,Capano:2019eae}. Since these tidal effects are significantly more pronounced at lower masses, detecting such a signal would provide great constraints on the EOS \citep{2024arXiv240517326K}.

X-Ray observations from the Neutron star Interior Composition Explorer (NICER) have provided important constraints on the EOS of neutron stars \citep{2019ApJ...887L..21R,2019ApJ...887L..24M}. However, these measurements currently rely on only a few well measured systems, leaving uncertainties particularly for SSM neutron stars. Gravitational-wave observations provide a unique opportunity to directly measure both the masses and tidal deformabilities of neutron stars, offering complementary information to  electromagnetic observations \citep{2021ApJ...918L..29R,2017Sci...358.1556C,2017ApJ...848L..30P,2019ApJ...887L..24M,2019ApJ...887L..21R,LIGOScientific:2017ync,Valenti_2017}, and further constrain the EOS across a wider mass range \citep{2024arXiv240517326K}.

Several searches have previously been performed for SSM binaries that span O1-O3 observing runs \citep{nitzsubsolar,nitzsubsolar1,nitzsubsolar2,ligosubsolar,2021arXiv210511449P}. Targeted searches for high mass ratio SSM binaries \citep{nitzsubsolarhighq} as well planetary sized objects have also been developed \citep{ssmcontinuous}. However, most prior modeled analyses have neglected tidal effects, treating compact objects as black holes. This omission leads to a significant loss in sensitivity to BNS signals, particularly in the low-mass regime, with up to a 78.4\% reduction in signal recovery for an equal-mass $0.2\,M_{\odot}$ \cite{2023PhRvD.107j3012B}. In practice, these estimates are optimistic as measures taken by searches to account for non-Gaussian noise will further reduce sensitivity \citep{Usman:2015kfa,2017PhRvD..95d2001M}.
To account for sensitivity loss, we developed a preliminary search that incorporated tidal deformability effects up to $\tilde\Lambda \sim 10^{4}$ \citep{2025ApJ...984...61K}; however, this did not encompass the the broad range of tidal deformabilities predicted for very low-mass neutron stars where a neutron star of $0.1~\m$ can have tidal deformabilities as high as $\sim10^7-10^8$~\citep{Capano:2019eae}. In this work, we reach up to tidal deformabilities of $\sim 7 \times 10^5$, allowing us to cover plausible EOSs for sources down to 0.25 \m for the first time; this has been enabled by the development of rapid search techniques~\cite{fir}.

We present the first results from a modeled SSM search sensitive to both BNSs and BBHs using data from the O4a observing run. Our most significant candidate has a false alarm rate of 5.73 per year. We constrain the merger rate of these systems with 90\% confidence to be $\mathcal{R}_{90}<2.0\times 10^5$ $\textrm{Gpc}^{-3} \textrm{yr}^{-1}$  and $\mathcal{R}_{90}<9.3\times 10^2$ $\textrm{Gpc}^{-3} \textrm{yr}^{-1}$ for a chirp mass of $0.1 ~\m$ and $0.7 ~\m$ respectively for both BNS and BBHs for O4a alone.  We observe more than double the improvement in the cumulative O1-O4a sensitivity in comparison to previous SSM searches \citep{nitzsubsolar,ligosubsolar}.

\section{Search} \label{sec:population}

We analyzed the gravitational-wave data using the \textsc{PyCBC} software framework \citep{Nitz:2017svb,Usman:2015kfa}, employing a matched-filter search designed to identify potential SSM signals. In this approach, the detector data are cross-correlated with a collection of modeled waveforms, known as a template bank, to measure the signal-to-noise ratio (SNR)~\cite{2012PhRvD..85l2006A}. Local maxima in the SNR, referred to as triggers, are examined for time and parameter coincidences across the detector network \citep{2012PhRvD..85l2006A}. A ranking statistic is assigned that distinguishes noise fluctuations from astrophysical candidates, and coincident triggers are assigned statistical significance through time-slide false alarm rate estimation~\citep{2017ApJ...849..118N,Davies:2020tsx}.

\begin{figure}[ht]
    \centering
    \includegraphics[width=\linewidth]{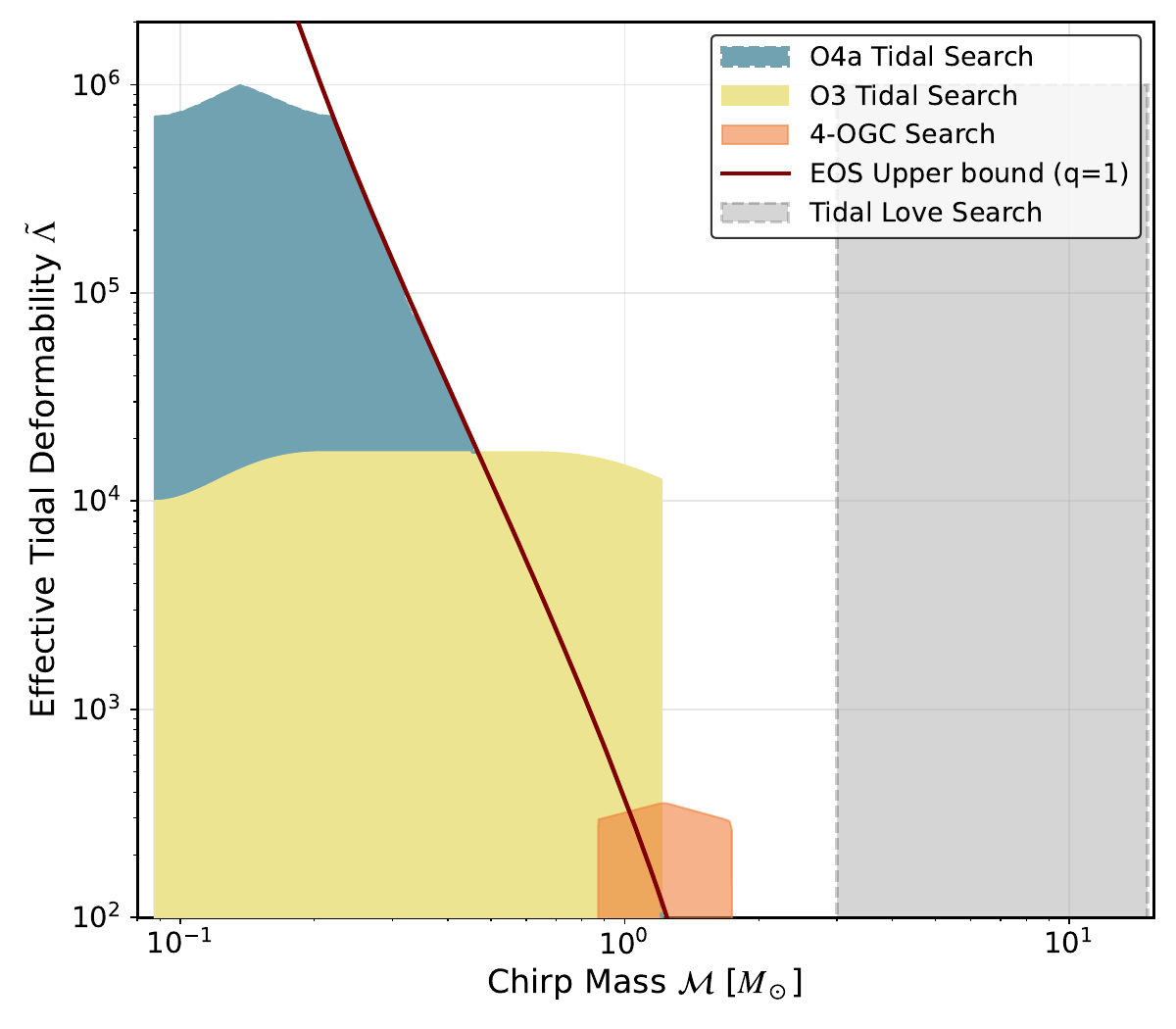}
    
    \vspace{0.3cm} 
    \begin{tabular}{|l|l|}
    \hline
    \textbf{Parameter} & \textbf{Range} \\ \hline
    Primary Mass \( m_1 \) & \( [0.1, 2] \, M_\odot \) \\ \hline
    Secondary Mass \( m_2 \) & \( [0.1, 1] \, M_\odot \) \\ \hline
    Aligned Spin \( \chi_1, \chi_2 \) & \( [-0.05, 0.05] \) \\ \hline
    Tidal Deformability \( \lambda_1, \lambda_2 \) & \( [0, 7\times10^5] \) \\ \hline
    \end{tabular}

    \caption{
    \textbf{Top:} The parameter space of searches as a function of the effective tidal deformability  $\tilde{\Lambda}$ \citep{2010PhRvD..81l3016H} and chirp mass $\mathcal{M}$ for previous tidal searches. The teal and yellows regions are the parameter space targeted in this work. Our previous O3 tidal search is depicted in yellow which directly overlaps which this search \citep{2025ApJ...984...61K}. The orange region indicates the range explored by the 4-OGC non-eccentric compact binary search \citep{Nitz:2021zwj}, and the gray region denotes the tidal Love number search \citep{Chia:2023tle}. The red line represents the soft EOS used as our analytical upper bounds on tidal deformability \citep{Capano:2019eae}. 
    \textbf{Bottom:} The parameter ranges spanned by our search, including component masses, spins, and tidal deformabilities. Waveforms were generated with a maximum duration of 512 seconds, a minimum frequency of 20 Hz, and a maximum frequency of 800 Hz. The bank was constructed such that 99.5\% of the templates recover at least $95\%$ of the optimal signal-to-noise ratio.
}
    \label{fig:parameter-space-table}
\end{figure}

Our template bank spans the parameter ranges shown in Figure ~\ref{fig:parameter-space-table}. 
We model isolated binaries using \textsc{TaylorF2} approximant \citep{Droz:1999qx,Blanchet:2002av,Faye:2012we}, and utilize stochastic placement methods \citep{2024ApJ...975..212K,2021PhRvD.104d3008H,2009PhRvD..80j4014H} to include tidal effects appropriate for low-mass compact binaries.
We improve upon our prior O3 search ($\lambda_{1,2} \leq 10^{4}$) by extending the search space towards more realistic values at very low masses, guided by a soft EOS model consistent with GW170817 from \citet{Capano:2019eae}. 
A complete extension down to $0.1$~\m would yield an impractically large bank ($\sim 10^{8}$ templates). For computational feasibility we imposed a maximum cutoff of $\lambda_{1,2} \leq 7\times10^{5}$, allowing for realistic EOS values for neutron stars down to $0.25~\m$. Our final bank contained approximately 25 million templates—roughly 15 million more than our previous search \citep{2025ApJ...984...61K}, making this search one of the most computationally expensive modeled search to date. 

Because the expected signals are long and computationally demanding to filter, we adopted the method used in \citet{nitzsubsolar}, which fixes the waveform duration to 512 seconds by adjusting the lower frequency cutoff upwards from a minimum of 20 Hz. Additionally, we set the upper frequency limit to 800 Hz. Tidal effects are most pronounced at high frequencies. We verified truncating the high frequency content did not lead to measurable SNR loss by calculating the overlap of our template against the full-bandwidth template.
To make this search feasible to conduct, we employ the ratio-filter de-chirping framework of \citep{fir,2025arXiv251112894M}, which enables efficient matched filtering of long-duration signals using compact finite-impulse-response filter representations. This method provided an eightfold improvement in template processing speed per core compared to previous analysis.

Previous SSM searches did not account for tidal effects but were sensitive to sub-solar black holes. The SSM search performed by \citet{ligosubsolar} used a template bank spanning $m_{1}\in[0.2,10]~\m,~m_{2}\in[0.2,1]~\m$ and spins up to 0.1 for masses less than $0.5~\m$. The SSM search by \citep{nitzsubsolar} employed a bank of about $7.8\times10^6$ templates, covering $m_1\in[0.1,7]~\m$, $m_2\in[0.1,1]~\m$, and eccentricities up to $e=0.3$. Our prior work was the first to incorporate tidal deformability into a SSM analysis, but did not cover the maximum tidal deformability predicted due to the computational cost associated with significantly expanding the template-bank volume \citep{2025ApJ...984...61K}. Our analysis is the first SSM search to incorporate tidal deformabilities up to $7\times10^5$ using the most sensitive gravitational wave data. 

\section{Constraining the Population of Sub-solar Binaries} \label{sec:results}
This search analyzed data from the O4a observing run of the Advanced LIGO detectors, spanning May 24, 2023 (1368195220 GPS) through January 16, 2024 (1389456018 GPS) \citep{Trovato:2019liz,GraceDB}. Virgo was not operating during this period; we only consider two-detector coincidences.  No confident detections were made, and among all potential candidates, we report the two most significant two-detector events in Table~\ref{tab:candidates}. 

\begin{table}[ht]
    \centering
    \caption{Top two-detector candidates, including intrinsic source parameters, merger time, and false alarm rate (FAR), and network SNR.}
    \label{tab:candidates}
    \begin{adjustbox}{width=\columnwidth} 
        \begin{tabular}{lccccc}
            \toprule
            GPS Time & $\mathcal{M}$ [$M_\odot$] & $\tilde{\Lambda}$ & FAR [$\text{yr}^{-1}$]  & Network SNR \\
            \midrule
            1369585385.539 & 0.32  & $2.29\times10^3$ & 5.73 & 8.98 \\
            1376009051.584 & 0.30  & $3.36\times10^3$ & 10.02 & 9.66 \\
            \bottomrule
        \end{tabular}
    \end{adjustbox}
\end{table}

\begin{figure}
    \includegraphics[width=\linewidth]{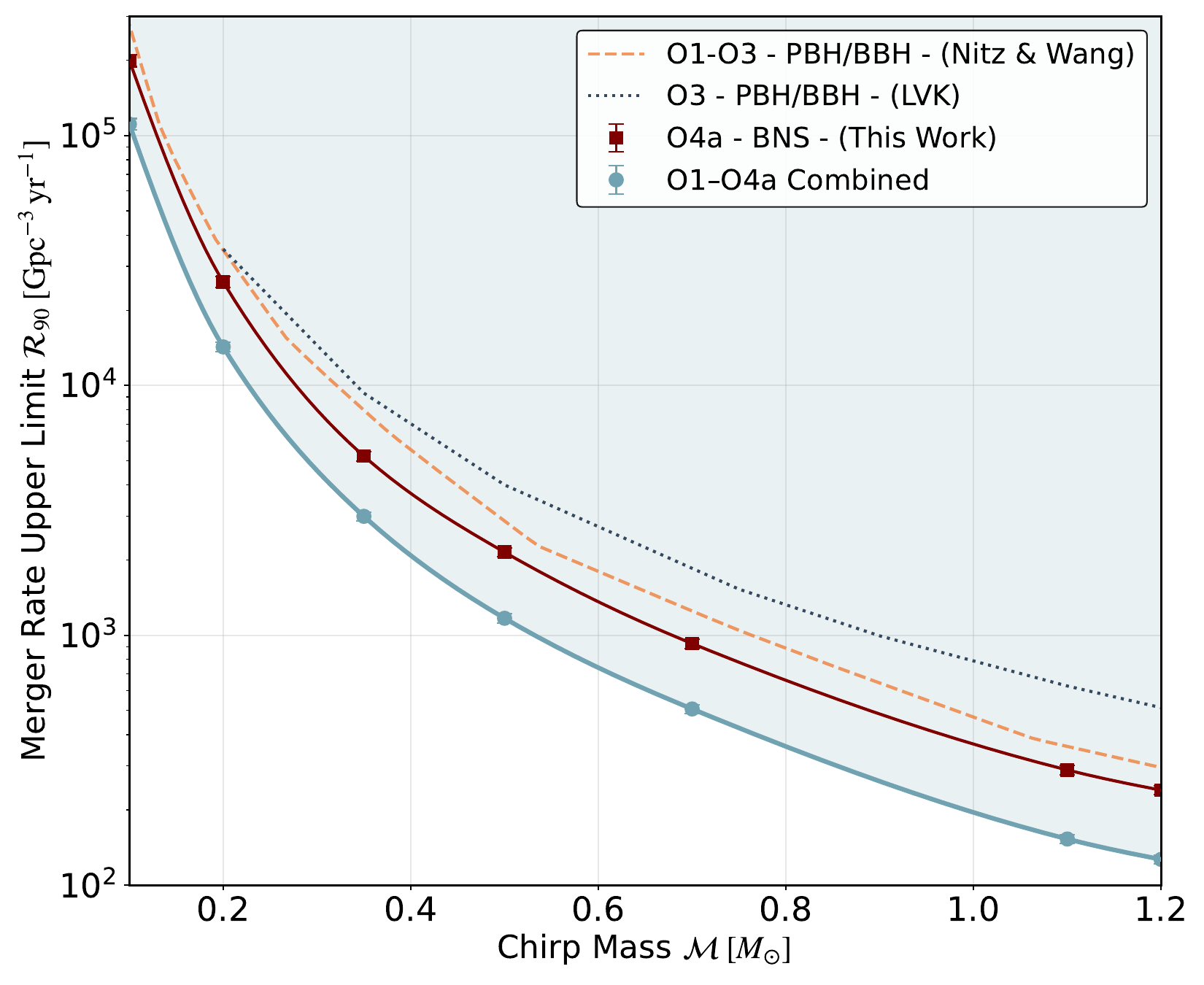}
    \caption{Upper limit of the merger rate at $90\%$ confidence $\mathcal{R}_{90}$ as a function of chirp mass $\mathcal{M}$. The red line represents the sensitivity for all tidal deformations covered by our search; we find that our search achieves equal sensitivity as a function of tidal deformability. The blue line represents the cumulative O1-O4a constraint on the black hole merger rate. Prior BBH constraints incorporating O1-O3 data \citep{nitzsubsolar}, and the latest O3 LVK sub-solar search  \citep{ligosubsolar} are depicted in the dashed orange line and dotted gray line respectively. For a $0.5 \, \m$ system, we observe a $1.9\times$ improvement in O4a alone compared to the LVK search. This is partly because the LVK search used a global lower-frequency cutoff of 45 Hz, which excludes signal power at earlier frequencies.
    }
    \label{fig:merger-rate}
\end{figure} 

Using a null detection, we can place 90\% confidence upper limits on the local merger rate $\mathcal{R}_{90}$ by calculating the sensitive volume through the recovery of simulated signals injected into the O4a data \citep{2009CQGra..26q5009B}. The fiducial signals were generated assuming an isotropic distribution in orientation and sky position and distances are distributed uniformly in volume. We sampled a grid of seven discrete chirp masses, $\mathcal{M}$ ranging from 0.1 up to 1.2 \m, and eight $\tilde{\Lambda}$ ranging from 0 up to $7\times10^{5}$. To maintain physical and search-consistency, we excluded injections that fall outside our search space; for instance, high-mass systems such as $\mathcal{M} = 1.2\,M_{\odot}$ were not paired with extreme tidal deformabilities like $\tilde{\Lambda} = 7\times10^{5}$, as these regions are excluded by our template bank's boundary.

Our rates are presented in Figure ~\ref{fig:merger-rate}. To construct the cumulative rate with the loudest event statistic \citep{2009CQGra..26q5009B}, we utilized the most significant candidate across all searches. The combined sensitivity through O4a exceeds the combined O1-O3 sensitivity by more than a factor of 2, reflecting the significantly improved performance of the detectors during this observing run \citep{2025PhRvD.111f2002C}.

\section{Constraints on the Primordial Black Hole DM Fraction}

\begin{figure}
    \includegraphics[width=\linewidth]{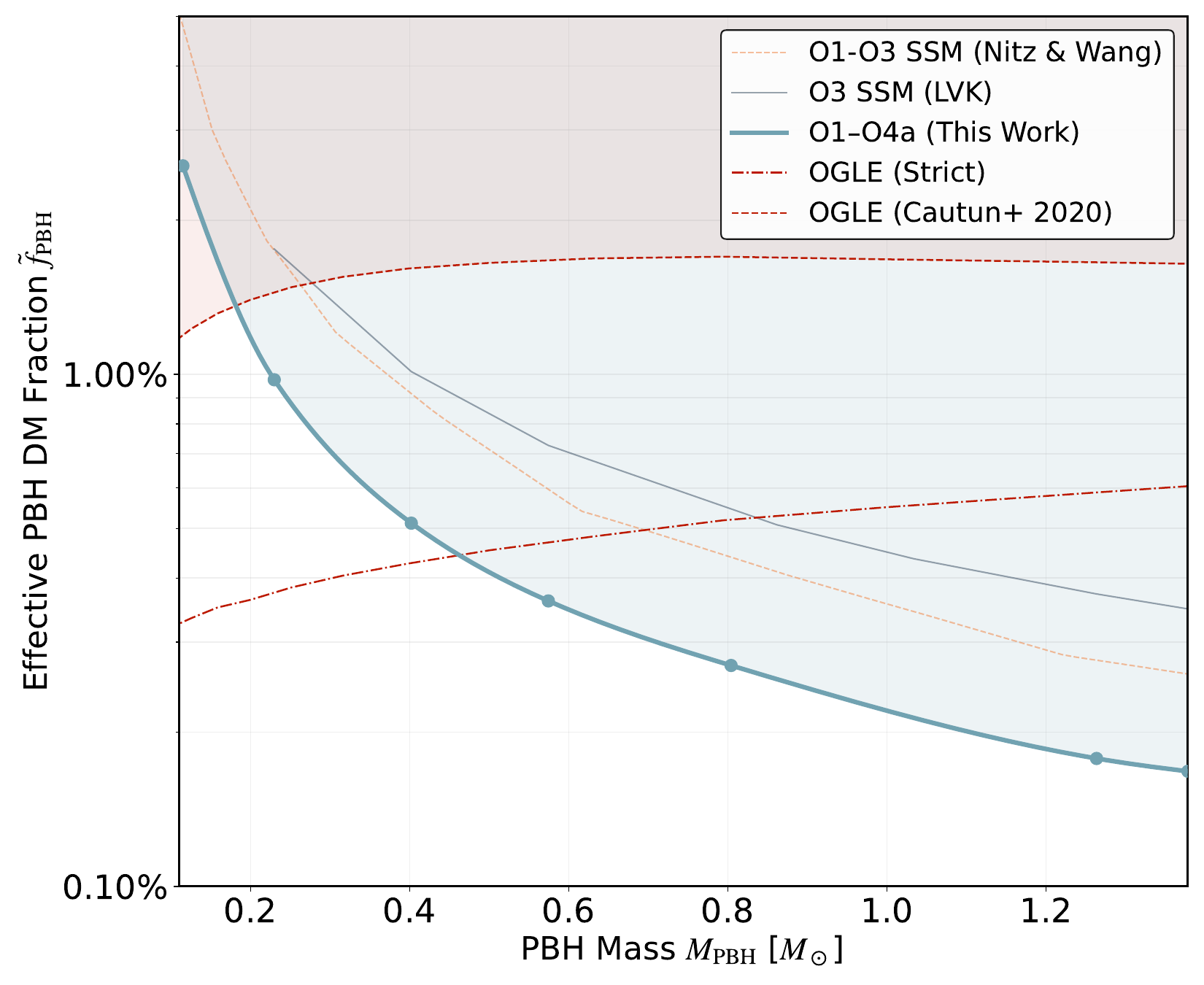}
    \caption{Constraints on the effective PBH DM fraction $\tilde{f}_{\rm PBH}$ as a function of mass $M_{\text{PBH}}$. The blue curve represents the cumulative constraints from O1-O4a. The orange line represents previous O1-O3 rates \citep{nitzsubsolar}, and the gray line represents the constrains from the O3 LVK SSM search \citep{ligosubsolar}. The red dashed curve and shaded region shows the OGLE microlensing limits obtained without imposing assumptions on the origin of the microlensing events, derived using the halo model of \citep{2020MNRAS.494.4291C}, and the red dotted dashed curved represents the constraints with the assumption that all observed events arise from stellar populations \citep{Mroz:2024mse}.}
    \label{fig:dmfrac}
\end{figure} 

Observed merger rate limits can be used to constrain the fraction of DM composed of PBHs. The fraction of DM is heavily dependent on the assumed population model for SSM Black Holes \citep{PhysRevD.79.023519,Kaplan:2009de,Feng:2009mn,Shandera:2018xkn}. We utilize the models described by \citet{2017PhRvD..96l3523A,2018ApJ...864...61C} as a fiducial reference which assumes that PBHs formed due to overdensities in the early Universe.

We represent these constraints in Figure \ref{fig:dmfrac}, where we assume a monochromatic mass distribution and can be applied to any arbitrary model. We utilize the effective fraction $\tilde{f}_{\text{PBH}}$, which scales the true mass fraction by a suppression factor that accounts for the reduction in the merger rate caused by disruptions of binaries after formation \citep{Raidal:2018bbj,Hutsi:2020sol}. For small rates $f_{\text{PBH}} \lesssim 0.1\%$, the suppression factor is not a predominant effect \citep{Raidal:2018bbj,Jedamzik:2020ypm,Hutsi:2020sol}. Our cumulative O1-O4a sensitivity places tight constraint at $0.4~\m$ PBH with $\tilde{f}_{\text{PBH}}<0.5\%$, $\sim1.8\times$ as tight of a constraint from previous works \citep{nitzsubsolar,ligosubsolar}.
We compare our results to the existing bounds from the OGLE microlensing survey \citep{Mroz:2024mse}. The strict limits assume that all observed microlensing events are due to known stellar populations. The relaxed limits make no assumptions about the origin of the events, allowing for a larger contribution from PBHs. Under the specific modeling assumptions adopted here, our inferred upper limits are $2-10$ times lower for for $M_\textrm{PBH}>0.25~\m$ for the relaxed model and $1.3-3.6$ times lower for $M_\textrm{PBH}>0.57~\m$ for the strict model.

\section{Discussions and Conclusion} \label{sec:conclusion}
A confirmed detection of a SSM compact object would have far-reaching implications for stellar astrophysics and fundamental physics. This observation would be difficult to reconcile with standard formation scenarios, which do not typically predict stable compact remnants with masses below $\sim 1 ~\m$ for neutron stars and black holes \citep{1983bhwd.book.....S}. Instead, it would point to alternative or exotic formation channels such as fragmentation processes in massive progenitor systems, other nonstandard mechanisms capable of producing SSM neutron stars, or the possibility that the object is of primordial origin in the case of a black hole \citep{Metzger_2024,greenPBHreview}.

In this work, we carried out a dedicated search for SSM binaries using the O4a from the Advanced LIGO detectors. We  incorporate tidal deformability as high as $7\times10^5$ in our waveforms, ensuring a robust search for BBH and highly deformable BNSs. Our best candidate was identified with a false-alarm rate of $5.73{~\textrm{yr}^{-1}}$. We derived upper bounds on the merger rate of systems, constraining it to $\mathcal{R}_{90}<1.1\times 10^5$ $\textrm{Gpc}^{-3} \textrm{yr}^{-1}$ for a $0.1~\m$ BNS or BBHs for the O1-O4a sensitivity, a $\sim 2.3\times$ improvement from previous O1-O3 rates. For a for a $0.8 ~\m$ PBH, we place new limits on the $\tilde{f}_\textrm{PBH}$ to be more than 1.6 lower than previous works, and a factor of 1.9 and 6.3 lower from the OGLE strict and relaxed model respectively.

This analysis underscores the computational challenges inherent to SSM searches. The long in-band duration and the inclusion of spin and tidal effects significantly expand the required template bank. While we extended the tidal range beyond previous work, covering the full physically allowed space (up to $\mathcal{O}(10^8)$ for $0.1~\m$) remains computationally prohibitive, as it would require tens of millions of additional templates. Utilizing the ratio-filter de-chirping framework \citep{fir,2025arXiv251112894M} made this search feasible, and achieved a $ 8\times$ improvement in computational performance, demonstrating that continued methodological advances will be essential for enabling more comprehensive SSM searches.

A wide range of search methodologies has been developed to detect gravitational waves from compact binaries, including fully-coherent matched-filter searches, semi-coherent analyses, and unmodeled burst searches \citep{Gadre:2018wsb,Privitera:2013xza,Sachdev:2019vvd,LIGOScientific:2019yhl,Klimenko:2015ypf,KAGRA:2021rmt}. Hierarchical search strategies have emerged as a promising avenue for future gravitational-wave analyses, offering a scalable approach to reduce computational cost while maintaining or improving detection sensitivity \citep{Gadre:2018wsb,ksoniheirch,rahulheirch,Mohanty:1997eu,Sengupta:2001gw}. 

Candidate observations such as the subthreshold trigger S250818k highlight the potential of multi-messenger data to probe the elusive SSM regime \citep{S250818k,ZTF25abjmnps,2022NatAs...6.1444D,Yuan:2024yyo,ligosubsolar,nitzsubsolar1,nitzsubsolar2,2025ApJ...984...61K}. However, the uncertain physical origin of the associated transient and the gravitational wave trigger suggest that the astrophysical interpretation remains tentative \citep{S250818k,ZTF25abjmnps}. If oriented optimally, a low chirp mass system like S250818k would lie near the horizon distance of the Advanced LIGO detectors, effectively at the limit of the detector sensitivity. Incorporating EOS informed tidal effects into future searches may potentially reveal SSM systems that remain just below present detection thresholds. S250818k occurred during O4c and future access to later data will allow a direct comparison. If SSM binaries arise from disk fragmentation~\citep{Piro:2006ja,Metzger_2024,Chen:2025uwd}, they may reside in dense gaseous environments where dynamical effects could modify the waveform. Such effects are not included in our analysis which may result in additional sensitivity loss.

If SSM systems are rare, next generation observatories may be necessary for their observations. Third-generation detectors such as the Einstein Telescope and Cosmic Explorer will extend the observable volume by orders of magnitude \citep{2010CQGra..27a5003H,2010CQGra..27s4002P,2019BAAS...51g..35R,Evans:2023euw}. For instance, a 40 km Cosmic Explorer is expected to reach as far as $z\sim100$, potentially detecting some of the early formed PBHs. The expanded reach of these detectors will enable detailed studies of the EOS across a broader mass spectrum, exploration of exotic compact objects, and potentially new avenues for probing the nature of DM.

\begin{acknowledgments} 

KK, KS, and AHN acknowledge support from NSF grant PHY-2309240. KK and AHN acknowledge the support from Syracuse University for providing the computational resources through the OrangeGrid High Throughput Computing (HTC) cluster supported by the NSF award ACI-1341006. KK acknowledges the support from the Open Science Grid (OSG) for providing additional computational resources to perform the search, which is supported by the National Science Foundation awards 2030508 and 2323298. Additionally, KK would like to express her gratitude to Rahul Dhurkunde, Ananya Bandopadhyay, Yifan Wang, and Labani Roy for all the discussions and feedback received. KK would like to thank Bradley J. Kavanagh for his efforts in maintaining and organizing the \textsc{PBHBounds} GitHub repository consisting of previous PBH constraints \citep{PBHbounds}, which greatly improved the comparative analysis in this work. Finally, KK would like to thank Tom Dent for pointing out a typo in the figure caption.

This research has made use of data or software obtained from the Gravitational Wave Open Science Center (gwosc.org), a service of the LIGO Scientific Collaboration, the Virgo Collaboration, and KAGRA. This material is based upon work supported by NSF's LIGO Laboratory which is a major facility fully funded by the National Science Foundation, as well as the Science and Technology Facilities Council (STFC) of the United Kingdom, the Max-Planck-Society (MPS), and the State of Niedersachsen/Germany for support of the construction of Advanced LIGO and construction and operation of the GEO600 detector. Additional support for Advanced LIGO was provided by the Australian Research Council. Virgo is funded, through the European Gravitational Observatory (EGO), by the French Centre National de Recherche Scientifique (CNRS), the Italian Istituto Nazionale di Fisica Nucleare (INFN) and the Dutch Nikhef, with contributions by institutions from Belgium, Germany, Greece, Hungary, Ireland, Japan, Monaco, Poland, Portugal, Spain. KAGRA is supported by Ministry of Education, Culture, Sports, Science and Technology (MEXT), Japan Society for the Promotion of Science (JSPS) in Japan; National Research Foundation (NRF) and Ministry of Science and ICT (MSIT) in Korea; Academia Sinica (AS) and National Science and Technology Council (NSTC) in Taiwan

\end{acknowledgments}

\section*{Data Availability}
All necessary files used in this work are publicly available in our GitHub repository \citep{github_o4a_sub}.

\bibliography{references}

\end{document}